\documentclass{article}
\usepackage{arxiv}
\pdfoutput=1

\usepackage[utf8]{inputenc} 
\usepackage[T1]{fontenc}    
\usepackage{hyperref}       
\usepackage{url}            
\usepackage{cite}
\usepackage{booktabs}       
\usepackage{amsfonts}       
\usepackage{nicefrac}       
\usepackage{microtype}      
\usepackage{lipsum}		
\usepackage{graphicx}
\usepackage{natbib}
\usepackage{doi}
\usepackage{threeparttable}
\usepackage{multirow} 
\usepackage{amsmath,amssymb,amsfonts}
\usepackage{algorithmic}
\usepackage{graphicx}
\usepackage{algorithm,algorithmic}
\usepackage{hyperref}
\usepackage{caption}
\usepackage{bm}
\usepackage{booktabs}
\usepackage{makecell}
\usepackage[inkscapeformat=png]{svg}
\usepackage[justification=centering]{caption}
\hypersetup{hidelinks=true}

\newcommand{\figref}[1]{Fig.~\ref{#1}}                      
\newcommand{\tabref}[1]{Table~\ref{#1}}                      
\newcommand{\eqnref}[1]{Eq. \eqref{#1}}                     %
\newcommand{\modelname}{WaveSleepNet}     

\usepackage{amsmath}

\title{WaveSleepNet: An Interpretable Network for Expert-like Sleep Staging}

\author{ \href{https://orcid.org/0009-0001-8431-0422}{\includegraphics[scale=0.06]{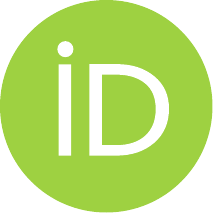}\hspace{1mm}Yan Pei} \\
	College of Biomedical Engineering \\
	Zhejiang University\\
	\texttt{summer\_bae@zju.edu.cn} \\
    \And
	{Jiahui Xu} \\
	Department of Neurology, Sir Run Run Shaw Hospital, School of Medicine \\
	Zhejiang University\\
	\texttt{xujiahui90@zju.edu.cn} \\
    \And
	{Feng Yu*} \\
	College of Biomedical Engineering \\
	Zhejiang University\\
	\texttt{osfengyu@zju.edu.cn} \\
    \And
	{Lisan Zhang*} \\
	Department of Neurology, Sir Run Run Shaw Hospital, School of Medicine \\
	Zhejiang University\\
	\texttt{zls09@zju.edu.cn} \\
    \And
	{Wei Luo*} \\
	College of Biomedical Engineering \\
	Zhejiang University\\
	\texttt{luo.wei@zju.edu.cn} \\
}

\renewcommand{\shorttitle}{\textit{ }}

\hypersetup{
pdftitle={WaveSleepNet: An Interpretable Network for Expert-like Sleep Staging},
pdfsubject={q-bio.NC, q-bio.QM},
pdfauthor={Yan Pei, and Wei Luo*},
pdfkeywords={Automatic sleep staging, interpretability, deep learning},
}

\begin{document}
\maketitle

\begin{abstract}
Although deep learning algorithms have proven their efficiency in automatic sleep staging, the widespread skepticism about their "black-box" nature has limited its clinical acceptance. In this study, we propose \modelname, an interpretable neural network for sleep staging that reasons in a similar way to sleep experts. In this network, we utilize the latent space representations generated during training to identify characteristic wave prototypes corresponding to different sleep stages. The feature representation of an input signal is segmented into patches within the latent space, each of which is compared against the learned wave prototypes. The proximity between these patches and the wave prototypes is quantified through scores, indicating the prototypes' presence and relative proportion within the signal. The scores are served as the decision-making criteria for final sleep staging. During training, an ensemble of loss functions is employed for the prototypes' diversity and robustness. Furthermore, the learned wave prototypes are visualized by analysing occlusion sensitivity.
The efficacy of WaveSleepNet is validated across three public datasets, achieving sleep staging performance that are on par with the state-of-the-art models when several \modelname s are combined into a larger network.
A detailed case study examined the decision-making process of the \modelname\ which aligns closely with American Academy of Sleep Medicine (AASM) manual guidelines. Another case study systematically explained the misidentified reason behind each sleep stage. \modelname’s transparent process provides specialists with direct access to the physiological significance of its criteria, allowing for future adaptation or enrichment by sleep experts.
\end{abstract}


\section{Introduction}
\label{sec:introduction}
The pervasive issue of sleep deprivation and the prevalence of sleep disorders represent a significant public health concern, with profound implications for morbidity and mortality. Sleep disturbances, which are closely associated with various psychiatric, neurodegenerative, and cardiovascular conditions, underscore the necessity for precise sleep analysis\cite{tobaldini:2018:short_sleep_cardiometabolic}. Polysomnography (PSG) is the cornerstone of sleep evaluation, holding a pivotal role in research and clinical practice by detecting unexpected events and thus revealing sleep abnormalities\cite{berry:2017:aasm}\cite{malhotra:2009:sleep_cardiovascular}.

The process of sleep staging, traditionally conducted by trained experts, involves extensive manual annotation, often necessitating several hours per patient. This labor-intensive approach incurs significant costs in terms of human resources, which can become a bottleneck for large-scale clinical research\cite{rosenberg:2013:sleep_medicine_inter_scorer}. Additionally, manual scoring is subject to inter-rater and intra-rater variability, posing challenges in maintaining consistency with scoring rules. 
Consequently, a succession of automatic sleep staging methodologies has been developed to support sleep specialists\cite{muto:2023:balance_visual_automatic}. These automated methods aim to reduce the subjectivity and resource demands of manual annotations while striving to match or surpass the accuracy of human experts.

The recent integration of deep learning (DL) methods into automatic sleep staging has demonstrated superior performance compared to traditional machine learning (ML) models, largely due to DL's ability to operate without extensive prior knowledge of physiological processes. Notably, Convolutional Neural Networks (CNNs) \cite{memar:2018:novel_cnn_sleep_stage}\cite{sars:2018:cnn_sleep_stage_single_channel}, Recurrent Neural Networks (RNNs) and and their hybrids \cite{sokolovsky:2019:dl_feature_sleep_stage}\cite{phan:2019:seqsleepnet} have been effectively employed in this field. However, the protracted training times associated with RNNs, due to their sequential data processing nature, and the critical need for contextual information in sleep signal analysis, have led to the adoption of transformer networks\cite{qu:2020:residual_attention_sleep_stage}\cite{phan:2022:sleeptransformer} and attention mechanisms\cite{edele:2021:attention_based_sleep_stage}.
These networks excel in capturing temporal dependencies. Among these developments, single-channel sleep staging has gained prominence for its user-friendliness and compatibility with home-based sleep monitoring.

Despite considerable advancements, the clinical adoption of automatic sleep staging algorithms remains limited. Discussions with leading sleep experts reveal a prevailing skepticism regarding the black box characteristic of deep learning models, a common concern within artificial intelligence in healthcare and medicine.  Both ML and DL require the extraction of representative features, which are then used to inform classification models.
ML methods\cite{hassan:2015:automatic_sleep_stage}\cite{saastamoinen:2006:computer_sleep_stage} offer greater interpretability but are limited by the labor-intensive process of manual feature engineering and often result in inferior performance in sleep stage classification when compared to DL methods. Conversely, DL algorithms have the advantage of autonomously learning features from raw data, without requiring prior knowledge of physiology. However, this advantage is a double-edged sword, as it obscures which specific sleep features are encoded. The consequent lack of explainability limits the adoption of deep learning models in clinical practice because clinicians often need to understand the reasons behind each classification to avoid data noises and unexpected biases\cite{ai-hussaini:2019:sleeper}. These issues highlight the necessity for advancements in DL-based sleep stage methodologies that prioritize interpretability to make the sleep stage process more explainable, by revealing more details of their inner workings.

Recent advancements have been made towards explainable sleep stage models. Dutt et. al\cite{dutt:2022:sleepxai} developed SleepXAI, an integrated CNN-CRF model that employs a modified Grad-CAM\cite{selvaraju:2017:grad_cam} for explainable multi-class sleep stage classification from single channel EEG signals.
This model highlights the signal segments most influential in stage prediction, enhancing explainability. Yet, the method falls short in explaining the reasoning process of how a network actually makes its decisions and clarifying the physiological relevance of its criteria. Phan et. al\cite{phan:2022:sleeptransformer} and Pradeepkumar et. al \cite{pradeepkumar:2022:explainable_sleep} both introduced transformer-based interpretable sleep stage models, a sequence-to-sequence model underpinned by transformer architecture, which interprets EEG data through self-attention scores visualized as heat maps to highlight sleep-relevant features captured from the input EEG signal. They mainly define the areas of focus without clarifying the relationship between the signal segments and their waveform characteristics, which can lead to ambiguity. Furthermore, these models concentrate on specific waveforms but neglect the significance of waveform prevalence in sleep staging task. For instance, alpha waves and theta waves are present in both Wake stage and N1 stage. However, the predominance of alpha waves typically indicates the Wake stage, while a higher proportion of theta waves is characteristic of the N1 stage—an important distinction that is not sufficiently captured by these models.

To address the above problem, we propose a waveform-sensing-based explainable neural network named \modelname. It enables network models to classify in a manner similar to human cognitive processes and explains their decision-making in a manner that agrees with the way sleep experts describe their own thinking in sleep staging tasks. How do human experts describe their process of sleep staging? According to AASM \cite{berry:2017:aasm} standards, which is briefly summarized in \tabref{tab:aasm}, sleep stages are classified by experts through the systematic analysis of characteristic waveforms (such as spindles, alpha waves, etc.). This process involves the examination of the presence and relative proportions of the specific waveforms within the input signals. To acquire such a sleep staging model that imitates this way of thinking and explains its reasoning process in a human-understandable way, we propose \modelname \ based on prototype learning\cite{bien:2011:prototype_selection}\cite{priebe:2003:class_cover_catch}\cite{chen:2019:this_looks_like} which is a classical form of case-based reasoning\cite{kolodner:1992:case-based-learning}. In the proposed model, a fully-convolutional neural network is utilized as a local feature extraction network to project raw EEG signals into a low-dimensional features. These deep features are segmented into patches, which are then compared with wave prototypes learned by the network within a latent feature space to calculate corresponding distances. These distances are fed into a subsequent module and transformed into scores which respectively reflect the presence of a wave pattern within the signal and the relative proportion of the wave pattern. These scores are then weighted and combined to derive the final sleep staging prediction. The wave prototypes are key elements of the \modelname\ model, signifying the latent representations of characteristic waveforms learned by the network that are capable of distinguishing between different sleep stages. To visualize the wave prototypes embedded in the network as signal waveforms, we employed occlusion sensitivity\cite{occlusion_study} analysis to equate each pattern with a signal segment derived from the training dataset. Furthermore, to better balance the accuracy and interpretability of the network, the proposed model incorporates an ensemble of loss functions, enabling the model to learn more diverse wave prototypes during the training process. In this way, our model is interpretable, in the sense that it has a transparent reasoning process when making predictions.

\setlength{\arrayrulewidth}{0.4mm}
\renewcommand{\arraystretch}{1.2}
\begin{table}[H]
\caption{Characteristics of each sleep stage according to the AASM rule}
\label{tab:aasm}
    \begin{center}
    \scalebox{0.5}{
    \resizebox{\linewidth}{!}{
        \begin{tabular}{c c}
        \hline
        Sleep Stage & Characteristics\\
        \hline
        Wake & \makecell[c]{$\bullet$ \text{alpha rhythm (8-13 hz)} \\$\bullet$ \text{ Eye blinks}\\ $\bullet$ \text{ Rapid eye movements (REM)}\\$\bullet$ \text{ Slow eye movements (SEM)}}\\
         \hline
         N1 & \makecell[c]{$\bullet$ \text{ More than 50\% of low-amplitude, mixed-frequency}\\ \text{ (LAMF) EEG activity (4-7 Hz)} \\ $\bullet$ \text{ Slow eye movements (SEM)} \\ $\bullet$ \text{ Vertex sharp waves (V waves)}}\\
         \hline
         N2 & \makecell[c]{$\bullet$ \text{ K-Complex}\\ $\bullet$ \text{ Sleep spindle (12-14 Hz)}}\\
         \hline
         N3 & \makecell[c]{$\bullet$ \text{ More than 20\% of Delta waves}\\ $\bullet$ \text{ Sleep spindle (12-14 Hz)}}\\
         \hline
         REM & \makecell[c]{$\bullet$ \text{ Rapid eye movements (REM)}\\ $\bullet$ \text{ Sawtooth waves (2-6Hz)}}\\
        \hline
        \end{tabular}
    }
    }
    \end{center}
\end{table}

The main contributions of this work are as follows:
\begin{itemize}
    \item[1)] 
     The proposed \modelname\ is a case-based model which automatically sleep scoring in a manner similar to the cognitive processes of sleep experts. To the best of our knowledge, this is the first application of case-based learning to the task of sleep staging. The effectiveness of \modelname\ model for sleep staging is validated on three public datasets.
     \item[2)] 
     Our model, \modelname, provides a level of interpretability characterized by a transparent reasoning process. It determines the presence and proportion of characteristic waveforms within a signal based on their proximity to the learned waveform prototypes, thereby classifying the sleep stage. Our experimental case studies also corroborate the alignment of our model's sleep staging logic with that of the AASM. 
     \item[3)] 
     We further analyzed the misclassification errors of the model which were not explored in prior studies. By systematically examining the reasons behind the incorrect classification of signals in various stages, we have provided direction for further improvements to the model to enhance the accuracy of sleep staging.
\end{itemize}


\section{METHOD}

\label{sec:method}
\begin{figure*}
    \centering
    \includegraphics[width=1.0\textwidth]{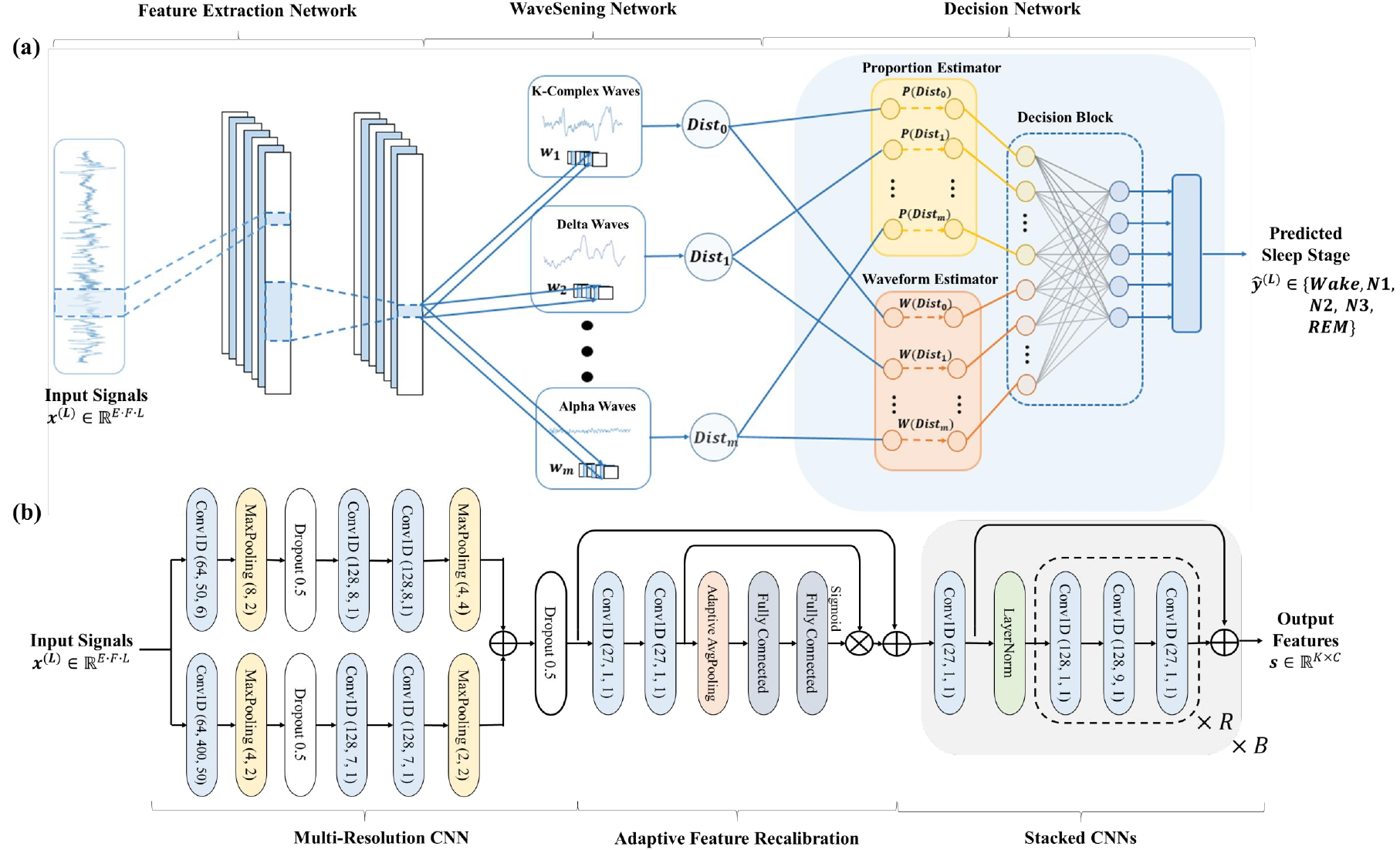}
    \caption{The architecture of the proposed \modelname. 
    (a) The overall model architecture. 
    (b) The structure of the Feature Extraction Network.}
    \label{fig:overall}
    \vspace{-14pt}
\end{figure*}

The proposed model is designed to classify an input sequence of single-channel EEG epochs $\boldsymbol{x}^{(L)}=(x_1, x_2, ..., x_L)$ into the sleep stage of the last input EEG epoch $\hat{\boldsymbol{y}}^{(L)}=(\hat{y}_L)$ (see \figref{fig:overall}). The input single-channel EEG signal is denoted as $\boldsymbol{x}^{(L)} \in \mathbb{R}^{E \cdot F\cdot L}$, where $E$ is the duration of an EEG epoch in seconds and $F$ is the sampling frequency. The predicted sleep stage for the $L$-th EEG epoch is presented as $\hat{y}^{(L)}\in \{Wake, N1, N2, N3, REM\}$. It follows the five-stage sleep classification in the AASM rule\cite{berry:2017:aasm}.

\figref{fig:overall} illustrates the overall framework of our \modelname\ model. It comprises three major components: Feature Extraction Network, WaveSensing Network and Decision Network. Firstly, the Feature Extraction Network derives deep features from the raw input signals, effectively capturing information across various temporal scales and channel dimensions. Subsequently, within the WaveSensing Network, the convolutional filters can be regarded as a series of specific waveform prototypes, which are constrained to be identical to some latent training patch. The process involves the comparison of extracted features against these learned waveform prototypes by calculating the $L^2$ distance for each waveform feature. These distances are subsequently fed into the Decision Network, which comprises two decision-making estimators. The two estimators converted the distances into scores to assess the presence and the relative proportion of the wave prototypes within the input signals based on the proximity to the wave prototypes. The scores are weighted combined to get the final prediction.

\subsection{Feature Extraction Network}
The role of the backbone network is to learn the feature map, $\mathcal{F}(\boldsymbol{x}^{(L)}):\boldsymbol{x}^{(L)}\in \mathbb{R}^{E \cdot F \cdot L} \mapsto s\in \mathbb{R}^{K\times C}$, in order to transform an input EEG epoch $\boldsymbol{x}^{(L)}$ into a high-level feature vector $s$ for representation. Herein, $K$ and $C$ denotes the dimensionality and the channel number of the extracted feature, respectively.

The architecture of the network is shown in \figref{fig:overall}(b), comprising modules of Multi-Resolution CNN (MRCNN), Adaptive Feature Recalibration (AFR) and stacked convolutional networks (stacked CNNs). The MRCNN and AFR are adopted from previous research\cite{supratak:2017:deepsleepnet}\cite{edele:2021:attention_based_sleep_stage}\cite{zhu:2023:masksleepnet}.

Given that different sleep stages are characterized by different frequency ranges,  we adopt the MRCNN to extract features in the signal across different scales. The MRCNN has two branches of CNNs with different kernel sizes, which are determined by the characteristics of the EEG signal and the signal sampling rate. In particular, Conv1D(64, 50, 6) in \figref{fig:overall} represents using a 1D convolution layer with 64 filters, a kernel size of 50 and a stride of 6. Similarly, MaxPooling(4, 4) refers to a MaxPooling layer with a kernel size of 4 and a stride of 4.

AFR is employed to model the inter-dependencies among features, enabling the selective emphasis on the most discriminative features via a residual Squeeze-and-Excitation (SE) block\cite{hu:2017:squeeze_excitation_network}. The features learned by the MRCNN are fed into two $1\times 1$ Conv1D convolutional layers, which integrate information across different channels and reduce the dimensionality of the features. Subsequently, a Squeeze-and-Excitation (SE) block is utilized to recalibrate the feature maps using “squeeze” operation of adaptive average pooling, followed by an “excitation” operation that uses two fully connected layers. 

To extract complex features in EEG signals, we adopted stacked CNNs that consists of $B $ stacked 1-D convolutional blocks, as shown in \figref{fig:overall}. 
Each convolutional block is formed by a convolutional layer, a residual connection, a 1-D Layer-Norm layer and $R$ unit convolutional layers which is formed by a sequence of 1-D convolutional layers. The residual path of a block serves as the input to the next block. LayerNorm refers to applying layer normalization \cite{jimmy:2016:layernorm}, stabilizing the hidden state dynamics in deep networks. In each convolutional unit, there’s a standard convolutional layer to increase feature dimensions, and depthwise separable convolution \cite{chollet:2017:depthwise_separable_cnn} which decouples the standard convolution operation into depthwise convolution and pointwise convolution to decrease the parameter numbers. 

\subsection{WaveSensing Network}

The WaveSensing Network architecture is inspired by ProtoPNet\cite{chen:2019:imag_prototype}, which dissects the image by finding prototypical parts and combines evidence from the prototypes to make a final classification. The WaveSensing Network capitalizes on trainable wave prototypes to transform the features learned by the Feature Extraction Network into a set of distance values between the features and the corresponding wave prototypes.

Given the deep features $s\in \mathbb{R}^{K \times C}$ extracted by the Feature Extraction Network, the network learns $M$ wave prototypes $W = \{ w_j\}_{j=1}^{M}$, whose shape is $K_1 \times C$ with $K_1\le K $. In our experiments, we set $K_1=1$. Since the channel ($C$) of each wave pattern is the same as that of the deep features but the length of each wave pattern is smaller than those of the deep features, each wave pattern will be used to represent some prototypical activation pattern in a patch of the deep features, which in turn will correspond to some prototypical waveform in the original EEG signals. Consequently, each wave pattern $w_j$ can be regarded as the latent representation of distinctive waveforms that are instrumental in differentiating sleep stages.  As a schematic illustration, the wave prototypes in \figref{fig:overall} corresponding to K-complex wave, Delta wave and Alpha wave.  

The deep features $s$ are segmented into $K$ one-point-overlapped patches which have the same shape as $w_j$. The distance $Dist_j\in \mathbb{R}^K$ between the $j$-th wave pattern $w_j$ and all patches of $s$ is calculated as \eqref{eq:dist}. The result is an activation map of distance scores whose value indicates the proximity between the wave pattern and the corresponding feature patches.

\begin{equation}
\label{eq:dist}
    Dist_j = Stack\{||\tilde{s} - w_j||_2^2 \thickspace for \;\tilde{s}\in patches(s)\}
\end{equation}

Every wave prototype $w_j$ is the latent representation of some training signal segments, which naturally and faithfully becomes the prototype's visualization. In this way, we can interpret the wave prototypes as visualizable prototypical signal segments. To determine which segment of the EEG signals from the training set $X_{Train}$ corresponds to wave pattern $w_j$, we initiate by selecting the EEG epoch $x_n$ that contains a feature patch $\tilde{s}_k$ with the minimum distance $Dist_{j}$ to $w_j$, as shown in \eqnref{eq:min_dist}. Subsequently, to pinpoint the relevant signal segment within $X_n$, we systematically occlude varying segments of the input signal with arrays of zeros, and monitoring the changes in $Dist_{j}$.The occlusion that results in a substantial increase in the distance $Dist_{j}$ identifies the signal segment that we then designate as the visual representation of $w_j$.

\begin{equation}
\begin{split}
\label{eq:min_dist}
    \min \{Dist_{j}\} &=\min_{s\in \mathcal{S}}\{||s-w_j||_2^2\},\thickspace \text{where} \thickspace \mathcal{S}= \\
    & \; \; \; \; \{\tilde{s}:\tilde{s}\in patches(\mathcal{F}(x_i)) \thickspace for \thickspace x_i \in x_{train} \} \\
    & = ||\tilde{s}_k-w_j||^2_2, \thickspace \text{where} \thickspace \tilde{s}_k\in patches(\mathcal{F}(x_n))
\end{split}
\end{equation}

\subsection{Decision Network}
Considering the fundamental scoring rationales for sleep staging, as outlined in \tabref{tab:aasm}, we propose a decision network composed of Waveform Estimator, Proportion Estimator, and Decision Block. The Waveform Estimator and the Proportion Estimator transform $Dist$, derived from WaveSensing Network, into scores that reflect the presence of a wave pattern within the signal and the relative proportion of the wave pattern, respectively. The Decision Block integrates these two scores to predict the final sleep stage accurately.

\subsubsection{Waveform Estimator}
The distance $Dist_j$ between deep features $s$ and the $j$-th wave pattern $w_j$ is converted into a similarity score as delineated by \eqnref{similarity}\cite{chen:2019:imag_prototype}. The similarity function employed is monotonically decreasing with respect to $Dsit_j$. The resultant activation map, which comprises similarity scores produced by wave pattern $w_j$, undergoes dimensional reduction via global max pooling to yield a single similarity score. This score quantitatively reflects the presence of a specific waveform within the input EEG epoch, as mathematically represented by \eqnref{wscore}. Consequently, a large output of  $WScore_j$, indicates that a deep feature patch closely aligns with the $j$-th wave pattern in terms of the L2-norm. This, in turn, suggests that a corresponding segment of the EEG signals has a similar concept to the characteristic waveform corresponding to the $j$-th wave pattern.

\begin{equation}
\label{similarity}
    Similarity_j = \log((Dist_j+1)/(Dist_j + \epsilon)
\end{equation}
                
\begin{equation}
\label{wscore}
    WScore_j = ReLU(BatchNorm(GlobalMaxPooling(Similarity_j) ))
\end{equation} 

\subsubsection{Proportion Estimator}
Similar to Waveform Estimator, the $Dist_{j}$ derived from WaveSensing Network is converted into a similarity score $Similarity_j$. Subsequently, this similarity score is transformed into a quantifiable measure that indicates the prevalence of the characteristic waveform within the input signals, which is shown in \eqnref{pscore}. In this score, we employ GlobalAveragePooling in place of the GlobalMaxPooling utilized in the Waveform Estimator. Thus, a large output of $PScore_j$ signifies not only the presence of a waveform in the input signal that closely resembles the $j$-th wave pattern but also indicates that such a waveform constitutes a significant proportion of the input signal.                
\begin{equation}
\label{pscore}
    PScore_j = ReLU(BatchNorm(GlobalAveragePooling(Similarity_j) ))
\end{equation} 
\subsubsection{Decision Block}

In the Decision Block, the outputs from Waveform Estimator ($WScore\in\mathbb{R}^M$), and the Proportion Estimator ($PScore\in\mathbb{R}^M)$, are concatenated to yield a composite score, denoted as $Score\in\mathbb{R}^{2M}$.  Next, a fully connected (FC) layer is utilized to assign suitable weights to the concatenated scores. The outputs of this layer are then fed into the sigmoid function to obtain the predicted sleep stage of the $L$-th epoch, denoted as $\hat{\boldsymbol{y}}^{(L)}$. This operation can be represented as shown in \eqnref{score}.             
\begin{equation}
\label{score}
    \hat{\boldsymbol{y}}^{(L)} = Sigmoid(FC(Concat([WScore, PScore])))
\end{equation} 

\subsection{Loss Function Ensemble}
The loss function of the network is carefully designed to optimize classification accuracy and enhance interpretability. In addition to the classification loss, there is a penalty on learned wave prototypes to promote their diversity and coverage of the data in latent representations. There are two error terms to reinforce the alignment of learned wave prototypes with distinct feature patches in the latent space. Furthermore, an L1 regularization term has been adopted to facilitate the wave prototypes selection for each sleep stage, which effectively improves interpretability.

We employ the standard cross-entropy loss to penalize the sleep stage misclassification. The cross-entropy loss is denoted by $L_{class}$, and is given by:          
\begin{equation}
\label{l_class}
    L_{class} = \frac {1}{n} \sum_{i=1}^n \sum_{c=1}^{N_C}-1[y_i=c]\log(P(\hat{y}=c))
\end{equation}
where $y_i $ is the label of the $i$-th EEG epoch and $P(\hat{y}=c)$ denotes the probability of the model predicting sleep stage class $c$. 

We use the average minimum squared $L_2$ distance between any two wave prototypes, $w_i$, $w_j$ to penalize the wave prototypes’ diversity error. The diversity loss, denoted by $L_{dist}$, is given by \eqnref{l_dist}. The logarithm function tapers large distances so that the penalty does not diminish or vanish, while the inclusion of the $\epsilon$ term enhances numerical stability.  By computing the inverse of the logarithm of the distances between wave prototypes, we penalize wave prototypes that are close in distance while making sure the minimum distance between wave prototypes does not become excessively large.              
\begin{equation}
\label{l_dist}
    L_{dist} = \frac{1}{\log(\frac{1}{m}\sum_{j=1}^m \min_{i>j\in[1, m]}||w_i-w_j||_2^2)+\epsilon}
\end{equation} 

There are two loss terms that relate the distance of the feature patches to the wave prototypes in latent space:
            
\begin{equation}
\label{l_r1}
    L_{R1} = \frac{1}{m}\sum_{j=1}^m\min_{i\in[1, n]}||w_j-s_i||_2^2
\end{equation} 
             
\begin{equation}
\label{l_r2}
    L_{R2} = \frac{1}{n}\sum_{i=1}^{n}\min_{j\in[1,m]}||s_i-w_j||_2^2
\end{equation} 

Here both terms represent the mean of the minimum squared distances. The minimization of $L_{R1}$ would necessitate each wave pattern to be as close as possible to at least one instance within the training set in the latent space, while the minimization of $L_{R2}$ would require every encoded training signal to be as close as possible to one of the wave prototypes. This balance ensures significant pixel-to-pixel correspondences between the wave prototypes and the training signals, thereby enhancing the representational fidelity of the model.

Furthermore, we have observed that without the constraints on the fully-connected layer within the Decision Block, the model indiscriminately assigns weights to all wave prototypes, regardless of their relevance to the specific sleep stage. This can lead to confusing model interpretability. To address this issue, we incorporated an $L1$ regularization term into our loss function to select significant wave prototypes in the sleep stage, effectively pushing the weights associated with less important wave prototypes toward zero.  The $L1 $ regularization term are shown as follows:
          
\begin{equation}
\label{l_1}
    L1=\sum_i|W_i|,\thickspace for\thickspace W_i\in W
\end{equation} 
where $W\in \mathbb{R}^{2M\times5}$ represents the weight matrices of fully-connected layer in the Decision Block.

Putting everything together, the overall objective function, denoted as $L$, can be formulated as:
\eqref{score}.    

\begin{equation}
\label{l_all}
    L = \lambda_{class}L_{class} + \lambda_{dist}L_{dist} + \lambda_{R1}L_{R1} + \lambda_{R2}L_{R2}+\lambda_{L1}L1
\end{equation} 
where $\lambda_{class}$, $\lambda_{dist}$, $\lambda_{R1}$, $\lambda_{R2}$ and $\lambda_{L1}$ are real-valued hyperparameters that adjust the ratios between terms.



\section{EXPERIMENTS AND RESULTS}

\subsection{Datasets}
To assess both the classification performance and the interpretability of our model, we conducted experiments on three public datasets, including SleepEDF-20, SleepEDF-78\cite{kemp:2000:sleepedfx} and Sleep Heart Health Study. In this study, we employed a single EEG channel from each dataset for classification.  For each dataset, we excluded EEG epochs with annotations unrelated to sleep stages, such as MOVEMENT class. Moreover, to address the five-class problem in datasets annotated with $R\&K$, we merged the N3 and N4 classes into a single N3 class according to the AASM criteria. \tabref{dataset} outlines the number of subjects, EEG channels used, evaluation scheme, number of reserved validation subjects, and sample distribution.

\setlength{\arrayrulewidth}{0.4mm}
\renewcommand{\arraystretch}{1.5}
\begin{table*}
\caption{Experiment settings and dataset statistics}
\label{dataset}
    \begin{center}
    \scalebox{0.8}{
    \renewcommand\arraystretch{1.5}
        \begin{tabular}{c c c c c c c c c c c }
        \hline
        \multirow{3}*{Dataset} & \multirow{3}*{No. of subjects} & \multirow{3}*{EEG channel} & \multicolumn{2}{c}{Experimenral setting} & \multicolumn{6}{c}{Class distribution} \\
        \cline{4-11}
        {} & {} & {} & \makecell[c]{Evaluation \\ scheme} & \makecell[c]{Held-out \\ validation set} & Wake & N1 & N2 & N3 & REM & Total \\
        \hline
        SleepEDF-20 & 20 & Fpz-Cz & 20-fold CV & 1 subject &\makecell[c]{8,285 \\ (19.6 \%)} & \makecell[c]{2,804 \\ (6.6 \%)} & \makecell[c]{17,799 \\ (42.1 \%)} & \makecell[c]{5,703 \\ (13.5 \%)} & \makecell[c]{7,717 \\ (18.2 \%)} & 42,308 \\
        SleepEDF-78 & 78 & Fpz-Cz & 10-fold CV & 7 subject &\makecell[c]{69,824 \\ (35.8 \%)} & \makecell[c]{21,522 \\ (10.8 \%)} & \makecell[c]{69,132 \\ (34.7 \%)} & \makecell[c]{13,039 \\ (6.5 \%)} & \makecell[c]{25,835 \\ (13.0 \%)} & 199,352\\
        SHHS & 5,793 & C4-A1 & \makecell[c]{Train/Test:\\0.7/0.3} & 100 subject &\makecell[c]{1,691,288 \\ (28.8 \%)} & \makecell[c]{217,583 \\ (3.7 \%)} & \makecell[c]{2,397,460 \\ (40.9 \%)} & \makecell[c]{739,403 \\ (12.6 \%)} & \makecell[c]{817,473 \\ (13.9 \%)} & 5,863,207\\
        \hline
        \end{tabular}
    \vspace{-14pt}}
    \end{center}
\end{table*}

\begin{figure*}
    \centering
    \includegraphics[width=0.8\textwidth]{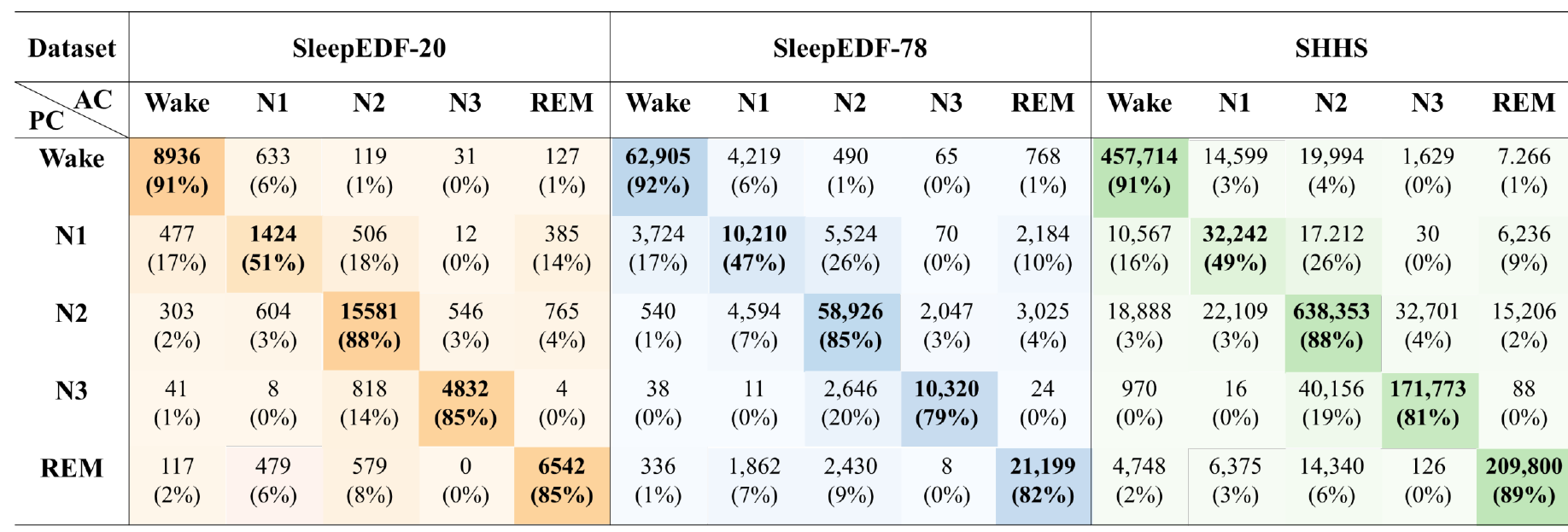}
    \caption{Confusion matrix of \modelname\ for SleepEDF-20, SleepEDF-78, and SHHS dataset. The values in parentheses indicate per-class recall. AC and PC denote Actual Class and Predicted Class, respectively.
}
    \label{confusion_matrix}
    \vspace{-14pt}
\end{figure*}

\subsubsection{SleepEDF-20}
This subset of the Sleep-EDF Expanded dataset (2013 version) was obtained from the PhysioBank\cite{goldberger:2000:physiobank} and included two studies. The Sleep Cassette (SC) study included 20 healthy Caucasian individuals aged from 25 to 34 years without any sleep-related medication. The Sleep Telemetry (ST) study included 22 subjects ranging in age from 18 to 79 years, aimed at investigating the effects of Temazepam on sleep. To avoid the influence of drug-related factors and following prior research, the SC subset was adopted in this study. The PSG recordings from the dataset were segmented into 30-second epochs and manually annotated by expert sleep clinicians according to the $R\&K$ manual. Each 30-second epoch was labeled as one of eight sleep stages (W, S1, S2, S3, S4, REM, MOVEMENT, UNKNOWN) by the sleep experts. Moreover, according to the common setting, we trimmed the PSG recordings to include a 30-minute window before and after the in-bed period.

\subsubsection{SleepEDF-78}
This dataset is an expanded version of the Sleep-EDF database, containing a total of 79 subjects ranging in age from 25 to 101 years, and comprising 153 complete overnight polysomnography (PSG) sleep recordings. The scoring standard and labeling methodology for the 30-s PSG epochs are the same as for SleepEDF-20. We evaluated our model using the Fpz-Cz EEG channel provided in the PSG recordings, which is sampled at a frequency of 100 Hz.

\setlength{\arrayrulewidth}{0.4mm}
\renewcommand{\arraystretch}{1.5}
\begin{table*}
\caption{Performance comparison between \modelname\ and state-of-the-art (SOTA) methods for automatic sleep scoring via deep learning; bold indicate the best model; the results indicated by * indicates the larger combined \modelname\ model.}
\label{metrics}
    \begin{center}
    \scalebox{0.8}{
        \begin{tabular}{p{1.5cm}<{\centering} c c c c c c c c c c c c}
        \toprule[2pt]
        \multirow{2}*{Dataset} & \multicolumn{3}{c}{Method} &  \multicolumn{3}{c}{Overall Metrics} & \multicolumn{5}{c}{Perclass F1 Score} \\
        \cline{2-12}
        {} & Model & Interpretable & Channel & ACC & MF1 & $\kappa$ & Wake & N1 & N2 & N3 & REM \\
        \hline
        \multirow{9}*{SleepEDF-20} & XSleepNet\cite{phan2021xsleepnet}& No &Single-Channel&86.3&80.6&0.81&-&-&-&-&- \\
        {}& TinySleepNet\cite{tinysleepnet}& No & Single-Channel&85.4&80.5&0.80&90.1&\textbf{51.4}&88.5&88.3&84.3 \\
        {}& SeqSleepNet\cite{phan:2019:seqsleepnet}& No&Multi-Channel&86.0&79.7&0.81&-&-&-&-&- \\
        {}& AttnSleep\cite{edele:2021:attention_based_sleep_stage}& No&Single-Channel&84.4&78.1&0.79&89.7&42.6&88.8&90.2&79.0\\
        {}&DeepSleepNet\cite{supratak:2017:deepsleepnet}& No&Single-Channel&82.0&76.9&0.76&-&-&-&-&-\\
        {}&SleepEEGNet\cite{sleepeegnet}& No&Single-Channel &84.3&79.7&0.79&89.2&52.2&\textbf{89.8}&85.1&85.0\\
        {}&SleepContextNet\cite{sleepcontextnet}& No&Single-Channel &84.8&79.8&0.79&89.6&50.5&88.4&\textbf{88.5}&85.0\\
        {}&Ours& Yes&Single-Channel&85.1&79.5&0.80&90.6&47.8&88.0&86.9&84.2\\
        {}&Ours*&Yes&Single-Channel&\textbf{86.5}&\textbf{81.0}&\textbf{0.87}&\textbf{92.2}&48.7&89.1&88.2&\textbf{86.7}\\
        \hline
        \multirow{10}*{SleepEDF-78}&XSleepNet\cite{phan2021xsleepnet}&No&Single-Channel&\textbf{83.9}&77.9&\textbf{0.78}&\textbf{93.3}&49.9&\textbf{86}&78.7&\textbf{84.2}\\
        {}&Korkalainen et al.\cite{Korkalainen2019}&No&Single-Channel &83.7&-&0.77&-&-&-&-&-\\
        {}&TinySleepNet\cite{tinysleepnet}&No&Single-Channel&83.1&78.1&0.77&92.8&\textbf{51.0}&85.3&81.1&80.3\\
        {}&SeqSleepNet\cite{phan:2019:seqsleepnet}&No&Multi-Channel&82.6&76.4&0.76&-&-&-&-&-\\
        {}&SleepTransformer\cite{phan:2022:sleeptransformer}&Yes&Single-Channel&81.4&74.3&0.74&91.7&40.4&84.3&77.9&77.2\\
        {}&U-Time\cite{utime}&No&Single-Channel&81.4&74.3&0.74&91.7&44.1&82.5&73.5&76.1\\
        {}&SleepEEGNet\cite{sleepeegnet}&No&Single-Channel&80.0&73.6&0.73&91.7&44.1&82.5&73.5&76.1\\{}&SleepContextNet\cite{sleepcontextnet}&No&Single-Channel&82.7&77.2&0.76&92.8&49.0&84.8&80.6&78.9\\
        {}&Ours&Yes&Single-Channel&82.5&77.1&0.76&92.5&47.5&84.7&80.8&80\\
        {}&Ours*&Yes&Single-Channel&\textbf{83.9}&\textbf{78.5}&0.77&\textbf{93.3}&49.8&85.9&\textbf{81.3}&82.2\\
        \hline
        \multirow{6}*{SHHS}& SleepTransformer\cite{phan:2022:sleeptransformer}&Yes&Single-Channel&\textbf{87.7}&80.1&0.82&92.2&46.1&88.3&\textbf{85.2}&\textbf{88.6}\\
        {}&XSLeepNet\cite{phan2021xsleepnet}&No&Single-Channel&87.6&\textbf{80.7}&0.82&92.0&\textbf{49.9}&88.3&85.2&88.6\\
        {}&IITNet\cite{iitnet}&No&Single-Channel&86.7&79.8&0.81&90.1&48.1&\textbf{88.4}&\textbf{85.2}&87.2\\
        {}&SeqSleepNet\cite{phan:2019:seqsleepnet}&No&Multi-Channel&86.5&78.5&0.81&-&-&-&-&-\\
        {}&Ours&Yes&Single-Channel&86.7 &79.1 &0.81 &92.1 &45.5 &87.6 &81.9 &88.5 \\
        {}&Ours*&Yes&Single-Channel&87.6 &\textbf{80.7} &\textbf{0.83} &\textbf{92.4} &49.5 &\textbf{88.4} &84.6 &88.5 \\
        \bottomrule[2pt]
        \end{tabular}
    \vspace{-14pt}}
    \end{center}
\end{table*}

\subsubsection{SHHS}
The SHHS dataset is a comprehensive multicenter cohort study designed to investigate the impact of sleep apnea on cardiovascular diseases. Initiated by Quan et al.\cite{quan:1997:shhs_dataset} and further explored by Zhang et al.\cite{zhang:2018:shhs_dataset}, the SHHS dataset encompasses two rounds of PSG recordings, labeled as Visit 1 (SHHS-1) and Visit 2 (SHHS-2). These recordings include a variety of physiological signals such as two-channel EEGs, two-channel EOGs, a single-channel EMG, a single-channel ECG, and two-channel respiratory inductance plethysmography. Sleep stages within each epoch of the recordings were classified according to the $R\&K$ rule into categories such as Wake (W), REM, N1, N2, N3, N4, MOVEMENT, and UNKNOWN. For the purposes of our study on single-channel EEG classification, we employed the C4-A1 EEG channel data in the experiment.

\subsection{Performance Metrics}
To evaluate the classification performance of different models, precision (PRE), recall (RE), and per-class F1 score (F1) were adopted as the performance criteria of each sleep stage, and the overall accuracy (ACC), macro F1-score (MF1), and Cohen’s Kappa (kappa) were adopted as the overall performance criteria. These metrics are calculated as follows:
\begin{equation}
    PRE=\frac{TP}{TP+FP}
\end{equation}
\begin{equation}
    RE=\frac{TP}{TP+FN}
\end{equation}
\begin{equation}
    F1 = \frac{2\times RE \times PRE}{RE + PRE}
\end{equation}
\begin{equation}
    ACC=\frac{TP+TN}{TP+FP+TN+FN}
\end{equation}

\begin{equation}
    MF1=\frac{1}{C}\sum_i^CF1_i
\end{equation}

\begin{equation}
    kappa=\frac{ACC-P_e}{1-P_e}=1-\frac{1-ACC}{1-P_e}
\end{equation}
where $TP, FP, TN, $ and $FN$ represent the number of true positives, false positives, true negatives and false negatives samples, respectively. $F1_i$ denotes per-class F1-score of the $i$-th class. $P_e$ represents the hypothetical probability of chance agreement.

\subsection{Implementation Details}
To evaluate the performance of \modelname\ model, we adopted a subject-wise k-fold cross-validation by dividing the subjects in each dataset (except SHHS) into k groups. As listed in \tabref{dataset}, we set k as 10, 20 for the SleepEDF-20, SleepEDF-78, respectively. For instance, subject-wise 20-fold cross-validation on SleepEDF-20 dataset with 20 subjects is thus leave-one-subject-out (LOSO) cross-validation. For each round, one group of subjects was used for testing and the remaining k-1 groups were divided into training and validation datasets.  The held-out validation set was used to track the validation loss for early termination in the training process. Unlike these datasets, the SHHS dataset was randomly divided in a ratio of 0.7 and 0.3 for training and testing, respectively. We further split 100 subjects of the training set into a validation set, as performed in \cite{phan2021xsleepnet}.

\begin{figure*}
    \centering
    \includegraphics[width=1.0\textwidth]{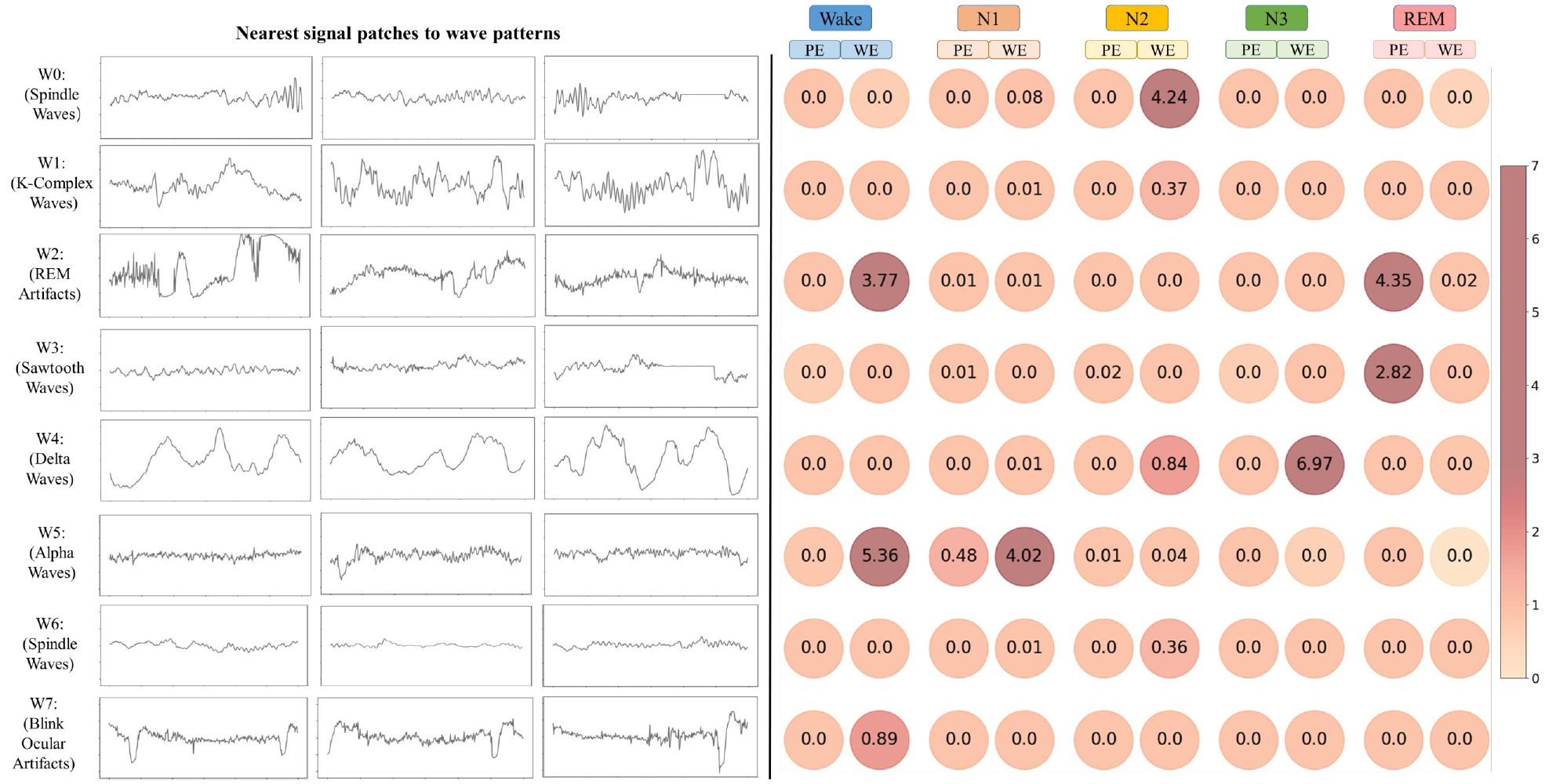}
    \caption{The reasoning process of the trained \modelname. The left part is the nearest signal patches in the training set to wave prototypes. The right part denotes the contribution of each wave pattern's proportion Estimator (PE) and waveform Estimator (WE) to various sleep stages. The wave prototypes are learned from the SleepEDF-78 dataset.
}
    \label{scoring}
    \vspace{-14pt}
\end{figure*}

We implemented out model using Pytorch 2.1.0, on a server equipped with a NVIDIA GeForce RTX 4090 GPU card. We adopted Adam optimizer with $\beta_1=0.9$, $\beta_2 = 0.999$. The initial learning rate is set to $5\times10^{-4}$, and the batch size is set to 64. The early-stop patience of the model training is set to 50. The number of EEG epochs for input signal $L$ is set to 10. The loss hyperparameters $\lambda_{class}, \lambda_{dist}, \lambda_{R1}, \lambda_{R2}, \lambda_{L1}$ are set to 50, 8, 9, 18 and 0.3, respectively.

\subsection{Sleep Staging Performance}

We present the sleep staging performance of \modelname\ through confusion matrixes depicted in \figref{confusion_matrix}. The confusion matrix is calculated by adding up all the scoring values of the testing data across all folds in each dataset. In the matrix, each row and column represent the classification results of sleep expert and \modelname, respectively. Notably, stage N1 achieves the lowest classification performance, where it is often misclassified as Wake and N2. We would further analyze the misclassification results in Section III.F.

Additionally, we compared the performance of \modelname\ model against various state-of-the-art approaches in terms of overall accuracy, macro F1-score, cohen kappa on three datasets. As shown in \tabref{metrics}, the sleep stage performance of \modelname\ model is comparable with that of the corresponding baseline (non-interpretable or interpretable) models. However, the accuracy of \modelname\ could be further improved by adding the logits of several \modelname\ models together.  Each \modelname\ model can be understood as a "scoring sheet" for sleep stages, as depicted in \figref{scoring}. Therefore, aggregating the logits from several \modelname\ models effectively creates the combined scoring sheet. In the combined model, $Score$s with wave prototypes from all these models are all taken into account to get the final predicted sleep stage. The combined model will maintain the same interpretable characteristic, though there will be more wave prototypes for each class. In this experiment, we combined five \modelname\ model together to get the final sleep stage performance. The test MF1 of combined \modelname\ can reach $81.0$, $78.5$ and $80.7$ for three datasets respectively, which is on par with the best-performing sleep stage model. Moreover, unlike other methods, \modelname\ has been simplified primarily to explore the feasibility of the proposed method, which only considers the local information in the input signals without taking into account the contextual information. This indicates that our model possesses potential for further improvement in staging performance, indicating a notable opportunity for advancement.

\subsection{Case Study I: Reasoning Process of \modelname}
In this section, we perform a case study to understand how our \modelname, which was trained on the SleepEDF-78 dataset, classifies sleep stages based on single-channel EEG signals. As shown in \figref{scoring}, the left part of the figure shows the nearest three signal segments in the entire training set to the wave prototypes encoded in the \modelname. It’s generally true that the nearest signal segments of a wave pattern all bear the same signal characteristic. After inspection by sleep experts, we identified the sleep characteristic waveforms that these signal segments belong to, that is, the sleep features corresponding to the wave prototypes learned by \modelname. In this case study, the wave prototypes learned by the network that can effectively stage sleep signals include spindle waves, K-complex waves, rapid eye movement (REM) artifacts, sawtooth waves, delta waves, alpha waves, and blink ocular artifacts. The right section of \figref{scoring} illustrates the contribution of each wave pattern relative to various sleep stages. The contribution is calculated by multiplying the $Score$ with the weights in the last FC layer. A higher contribution value not only indicates the presence or greater proportion of a sleep feature within the input signal but also carries a greater weight in determining the sleep stage. Notably, we found that while N1 stage is determined by W5 waveforms that are referenced as alpha waves, its $Score$ value in W5 was lower than that of Wake stage signals. This is because its characteristic waveform does not exactly match alpha waves, but the LAMF waves that are similar to alpha waves but with a slower frequency of at least 1Hz in comparison.

\begin{figure}
    \centering
    \includegraphics[width=0.5\textwidth]{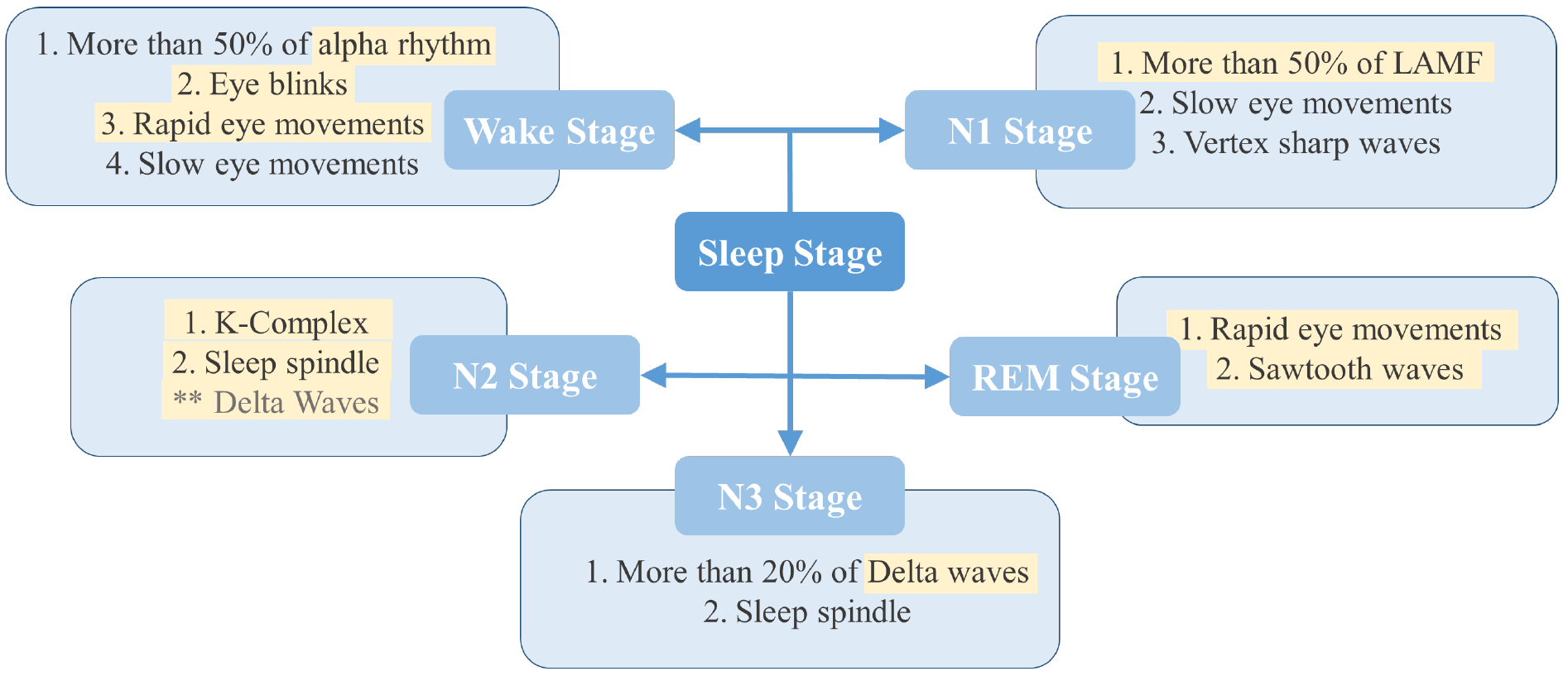}
    \caption{Comparison of the Classification Logic in \modelname\ Model versus the AASM Manual Guidelines for Sleep Staging. The numbered annotations indicate the classification logic according to the AASM Manual, light yellow boxes highlight the classification logic implemented by the \modelname\ model. ** denotes the classification rules unique to \modelname\ and not illustrated in AASM guidelines.
}
    \label{aasm}
    \vspace{-14pt}
\end{figure}

Moreover, the reasoning process of \modelname\ for sleep staging can be outlined in \figref{aasm}. The results in \figref{aasm} show that \modelname\ produces promising interpretable outcomes. Its decision-making process significantly overlaps with that of human manual annotators\cite{berry:2017:aasm}. Notably, we observed that \modelname\ seems to consider Delta waves (W4) when predicting the sleep N2 stage, a deviation from the staging criteria outlined in the AASM manual. This is because K-complex waves are considered Delta waves if they fit the definition of slow activity, which is mentioned in the AASM.

\begin{figure}
    \centering
    \includegraphics[width=0.5\textwidth]{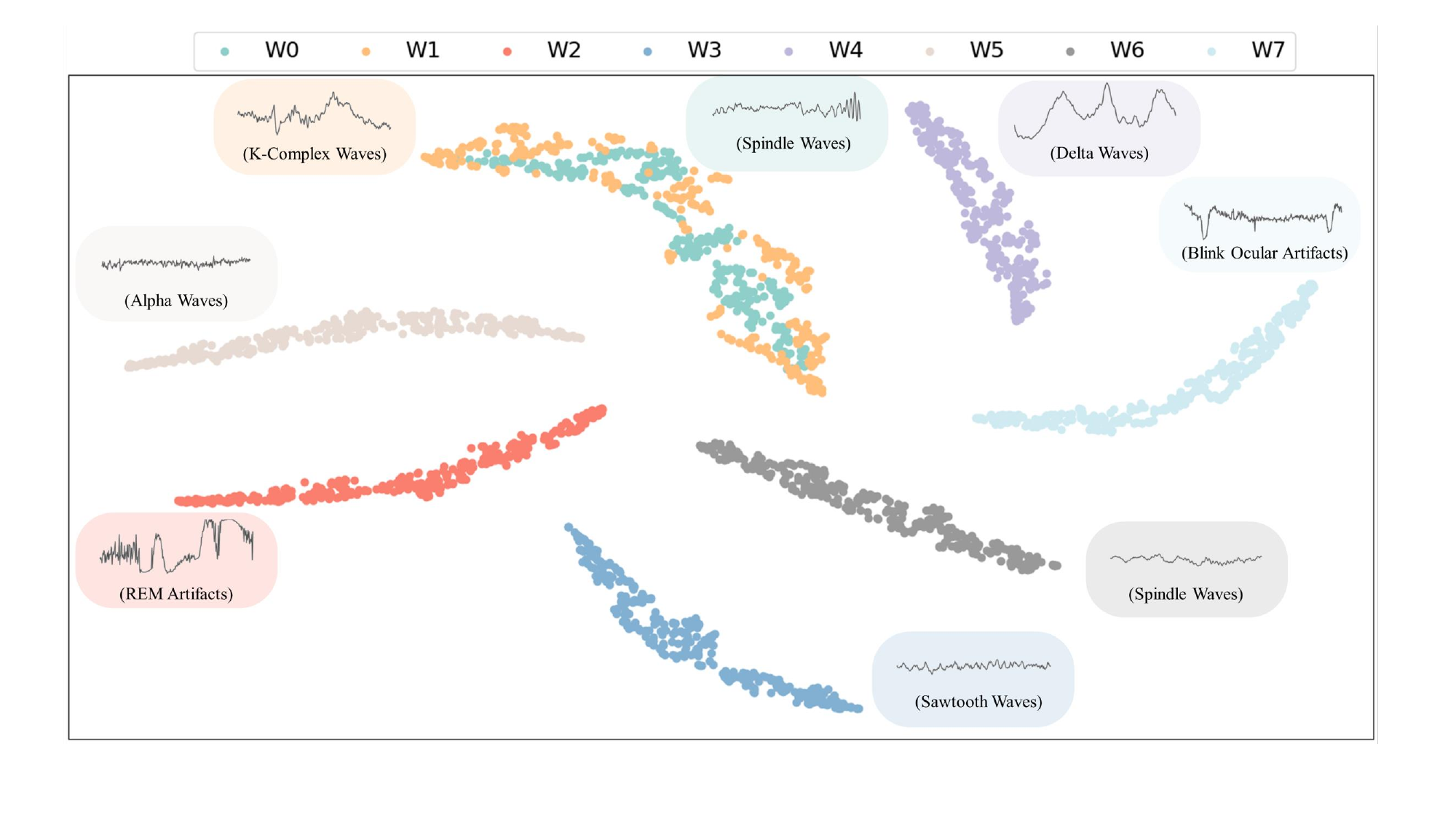}
    \caption{Learned wave prototypes of \modelname\ trained in the SleepEDF-78 dataset
}
    \label{scatter}
    \vspace{-14pt}
\end{figure}

Additionally, we examine $Score$s, as the latent features of each wave pattern, via t-SNE projections where proximity in 2-D space suggests that points are “close” in distance in the original latent space, shown as \figref{scatter}. The $Score$s of each wave pattern are extracted from a series of EEG signals which form the closest patch-wave pattern pair in the latent space with the corresponding wave pattern. This shows that the learned wave prototypes are easily separated and well-structured. However, it was observed that W0 (Spindle Waves) and W1 (K-Complex Waves) are proximal yet not entirely congruent within the 2-D projection space. This phenomenon can be attributed to the consistent interaction between K-Complex waves and sleep spindles, as spindles frequently coincide with K-Complex waves\cite{kokkinos:2010:k-complex_spindle}. The two wave prototypes frequently exhibit overlapping characteristics, leading to their relative proximity within the feature space.
Furthermore, we found that despite W0 and W6 are all represent spindle characteristic waves, they do not exhibit proximity in the 2-D space. This indicates that, although they share similar frequencies, they represent distinct categories of spindle waves. Notably, W6, in comparison to W0, demonstrated a lower amplitude, suggesting that W6 might be the spindle wave interferences from other EEG channels.

\subsection{Case Study II: Misclassification Analysis of \modelname}

\begin{figure}
    \centering
    \includegraphics[width=0.5\textwidth]{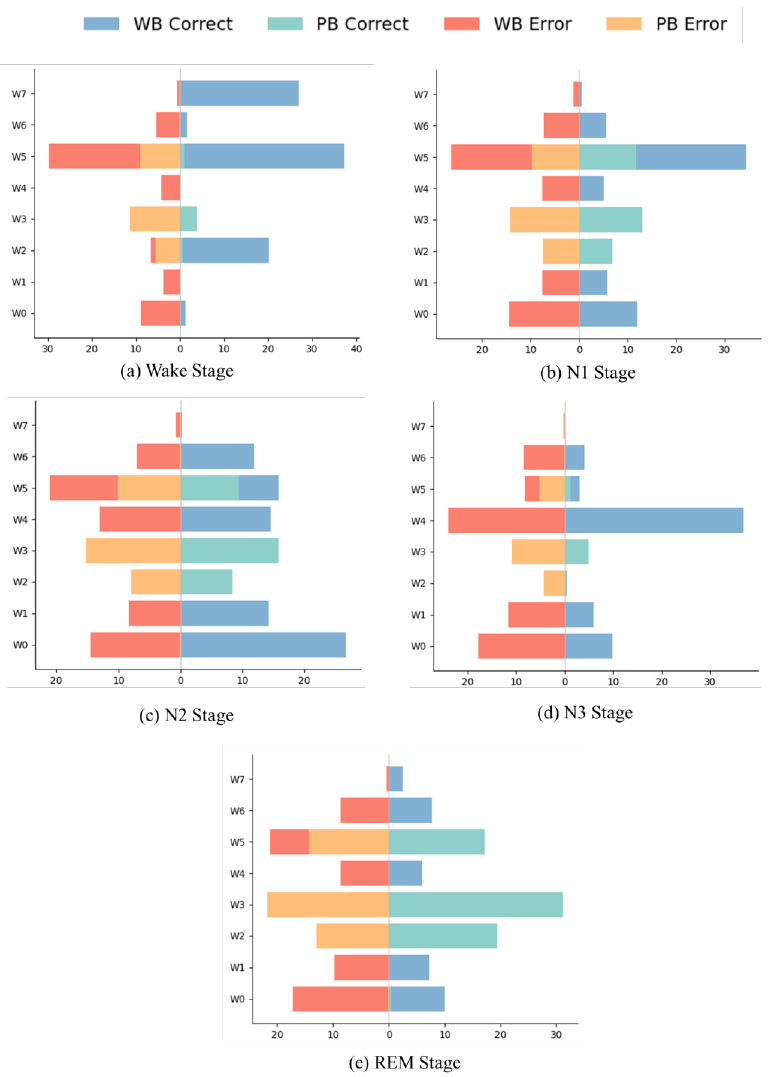}
    \caption{The comparison of $Score$ values for accuractely classified instances and mistakenly classified instances. $WE\enspace Correct$ and $WE\enspace Error$ represent the $WScore$ for correctly and incorrectly classified instances, respectively, while $PE\enspace Correct$ and $PE\enspace Error$ denote the $PScore$ for accurately and mistakenly classified instances, respectively.
}
    \label{error_score}
    \vspace{-14pt}
\end{figure}

As an explainable model, \modelname\ not only provides a reliable reasoning process for accurate predictions, but it also explains its misclassifications. In this section, we examine the reasons behind \modelname's mispredictions across various sleep stages. We achieve this by extracting the $Score$ values of wave prototypes. As illustrated in \figref{error_score}, we conducted a comparison between the $Score$ values of accurately classified instances and that of mistakenly classified instances. Fro
m \figref{error_score}, we can see that \modelname \ determines sleep stages by identifying specific waveforms in EEG signals. Therefore, if the characteristic waveforms of a certain sleep stage signals weaken or are not significant enough, they may be incorrectly classified into other stages. We can analyze each stage one by one:

\textbf{1) Wake Stage:} The waveforms used to determine the Wake stage are Blink Ocular Artifacts (W7), REM artifacts (W2), and alpha waves (W5), while the waveforms used to determine the N1 stage are alpha waves (W5). When the Blink ocular artifacts and REM artifacts in the input signals of the Wake stage are not pronounced, the signal may be misidentified as N1 stage, as indicated by the results of the confusion matrix.

\textbf{2) N1 Stage:} The classification of N1 sleep stage depends on the $Score$ value of alpha waves (W5), while the classification of N2 sleep stage mainly depends on the $Score$ value of spindle waves (W0). When the frequency of alpha waves appearing in stage N1 signals is similar to that of spindle waves, the W0 $Score$ value will increase and thus the signal will be mistakenly classified as stage N2.

\textbf{3) N2 Stage:} Spindle waves (12-14 Hz) and alpha waves (8-13 Hz) are similar in frequency, leading to some spindle waves being incorrectly identified as alpha waves. If there is no K-complex wave (W1) present in the signal at this time, this may result in the input signal being erroneously classified as N1 sleep stage. In the N2 stage, V-waves are similar to REM artifacts (sharply contoured waves with a duration <0.5 seconds), thus some V-waves may be mistakenly identified as REM artifacts (W2). Furthermore, both N2 and REM stages have sawtooth waves (W3). These lead to some signals being misclassified as REM stage.

\textbf{4) N3 Stage:} The waveforms used to determine the N3 stage is Delta Waves (W4). When there are prominent spindle waves (W0, W6) in the input signal and Delta waves (W4) are not significant, the signal may be mistakenly identified as N2 stage.

\textbf{5) REM Stage:} The similarity between low-frequency alpha waves during REM stage and alpha waves (W5) causes a high W5 $Score$ value, making it easy for the input signal to be misclassified as N1 sleep. Additionally, the low-frequency alpha waves are similar to spindle waves (W0, W6), and if sawtooth waves (W3) are present in the signal simultaneously, it may be misclassified as N2 stage.

\section{Discussion}
In this paper, we propose an interpretable sleep staging model, \modelname, that explains decision-making process in a manner that aligns with the cognitive processes of sleep experts. By conducting experiments on different datasets, we demonstrated its comparable sleep staging performance with the existing state-of-the-art methods. Additionally, a case study was conducted to discuss the reasoning process of \modelname\ in the task of sleep staging. The results demonstrate that it yields highly interpretable outcomes, which align closely with the logic used by human manual annotators. Beyond this, the study also delves into the reasons behind classification errors made by \modelname\ during the sleep staging process. This investigation offers valuable insights for future improvements to sleep staging methods. 

From Section III.E, we discovered that ocular artifacts play a significant role in the model’s classification process. This finding is not limited to the specific case study, it has been observed across all datasets. This indicates that, in the task of sleep staging using single-channel EEG signals, ocular movement features are still encapsulated within the single-channel EEG signals in the form of artifacts. This might explain why many single-channel sleep staging models perform comparably to multi-channel models\cite{phan2021xsleepnet}\cite{phan:2022:sleeptransformer}\cite{sleepcontextnet}\cite{edele:2021:attention_based_sleep_stage}\cite{wei:2024:contex_atten_sleep}. It also clarifies why models utilizing the Fpz-Cz channel often achieve better staging outcomes than those using Pz-Oz channel\cite{edele:2021:attention_based_sleep_stage}\cite{dutt:2022:sleepxai}\cite{huang:2022:senet}, as the Fpz-Cz channel is more susceptible to ocular interference. Nevertheless, ocular artifacts can also disrupt the staging process. As illustrated in section III.F, the similarity between V-waves and REM ocular artifacts can lead to the misclassification of N2 stage signals as REM stage. Furthermore, accurate identification of EOG features typically requires the collaborative judgment of signals from two EOG channels, and relying solely on single-channel ocular artifacts could result in misinterpretation of these features. Therefore, future work should aim to accurately leverage ocular artifacts in single-channel EEG sleep staging without introducing interference.

Through the analysis of misclassifications by \modelname \ in section III.F, we noted that the most significant issue arises from the confusion between the alpha waves (8-13Hz) during the Wake (W) stage, the attenuated alpha rhythm waves in the N1 stage, the spindle waves in the N2 stage (11-16Hz), and the low-frequency alpha waves during the REM stage. This confusion is primarily due to the overlapping frequencies of these waves, making it challenging to accurately distinguish them within the signal. Misidentification can lead to incorrect sleep staging. This problem is not unique to \modelname; it is a common challenge faced by all neural network models when processing single-channel EEG signals. In future work, enhancing the separation of these waveforms through advanced signal processing or detection techniques could potentially improve the accuracy of sleep stage classification models.

The annotators designated by sleep experts should not be regarded as the absolute gold standard, as there is a substantial inter-expert variability of 65-85\% \cite{norman:2000:interobserver_sleep_scores}. Therefore, investigating the reasoning process of the model in sleep staging may be more valuable than determining the accuracy of the staging itself. The international consensus criteria are the experiences or theories derived from sleep researchers' extensive work over the years, and they are constantly being updated and revised. We don’t have to make them an absolute ground truth to be essentialized \cite{muto:2023:balance_visual_automatic}. Our \modelname\ model, as a data-driven method, identifies discernible features and rules from a large number of sleep datasets, facilitating sleep stage differentiation. This, in turn, could be extremely interesting in revealing new sleep signatures, but their relevance still need to be validated by human experts.

Our model has been simplified primarily to mainly explore the feasibility of the proposed method. Therefore, further adjustments to the base model could be explored in future work to enhance the model’s interpretability and sleep staging performance.

Firstly, extending the ability of the network would unquestionably enhance the classification performance. In this study, a convolutional model was utilized for local feature extraction from input signals, but the contextual relationships between epochs were not considered. Future research should explore integrating memory-augmented neural networks like RNNs and LSTMs into the \modelname\ to maintain interpretability while learning epoch contextual relationships, thereby enhancing sleep staging accuracy. Additionally, our Feature Extraction Network in \modelname\ adapted a simplified CNN model inspired by Attnsleep\cite{edele:2021:attention_based_sleep_stage}. Future efforts should aim to investigate more advanced network architectures for improved extraction of sleep waveform features.

Second, the \modelname\ achieves sleep staging by combining waveform and proportion Estimators, mirroring the methodology of sleep experts. However, the decision-making analysis of the trained model reveals discrepancies with the AASM guidelines in the proportion-based reasoning process. The divergence may be attributed to the neural network’s extensive feature space, making it challenging for the network to identify the global optimum solution that aligns perfectly with human expert annotators. To make the staging process more aligned with human expert logic, future research could consider further exploration of integrating expert knowledge into the model.

\section{Conclusion}

Interpretability and accuracy represent a trade-off in machine learning modeling. Especially in the age of deep learning, many accurate models are black-box models that lack transparency in explaining their predictions, limiting their clinical acceptance. In this paper, we present a deep learning model \modelname\ that provides accurate predictions as well as offers interpretability by revealing more details of its inner workings. The model was developed and evaluated using the SleepEDF-20, SleepEDF-78, and SHHS datasets, demonstrating performance in sleep staging that is comparable to state-of-the-art models. Moreover, a combined larger \modelname\ model outperforms state-of-the-art methods under various evaluation metrics. A qualitative case study was conducted to provide explanations through an interpretable decision-making process that closely align with AASM. 
Furthermore, we innovatively conducted an analysis of the model's misclassifications, systematically explaining the reasons for each misidentified stage, which helps to better understand and address classification degradation. The added interpretability in our model allows the experts to have direct access to the physiological meaning of the criteria within the model, indicating a possibility for human-machine collaboration for sleep stage classification. 

\bibliographystyle{unsrtnat}

\end{document}